\documentclass[%
 twocolumn,
 amsmath, amssymb,
 aps,
 prl,
 longbibliography
]{revtex4-1}

\usepackage{graphicx}
\usepackage{dcolumn}
\usepackage{booktabs,subcaption,amsfonts,dcolumn}
\usepackage{multirow}
\usepackage{diagbox}

\usepackage{graphicx}
\usepackage[english]{babel}
\usepackage{color}   
\usepackage{bbm}
\usepackage{tabularx}
\usepackage{algorithm}
\usepackage{algpseudocode}
\usepackage{bm}
\usepackage{tikz}
\usepackage[pdfpagelabels,plainpages=false,bookmarks=true,colorlinks,linkcolor=red,urlcolor=blue,citecolor=blue]{hyperref}

\newcommand{\Ising}{\text{Ising}}
\newcommand{\TFIC}{\text{TFIC}}
\newcommand{\Potts}{\text{Potts}}
\newcommand{\DE}{\text{DE}}
\newcommand{\GGE}{\text{GGE}}

\DeclareMathOperator{\Tr}{Tr}

\renewcommand{\vec}[1]{{\bf #1}}
\newcommand{\braket}[1]{\langle #1  \rangle}
\newcommand{\ket}[1]{| #1  \rangle}
\newcommand{\bra}[1]{\langle #1|}

\newcommand{\norm}[1]{\lVert #1 \rVert}
\newcommand{\GS}{\text{GS}}

\begin{document}

\title{Dissipative dynamics in isolated quantum spin chains after a local quench}

\author{Yantao Wu}

\affiliation{
The Department of Physics, Princeton University
}
\date{\today}
\begin{abstract}
We provide numerical evidence that after a local quench in an isolated infinite quantum spin chain, the quantum state locally relaxes to the ground state of the post-quenched Hamiltonian, i.e. dissipates. 
This is a consequence of the unitary quantum dynamics.  
A mechanism similar to the eigenstate thermalization hypothesis is shown to be responsible for the dissipation observed.
We also demonstrate that integrability obstructs dissipation. 
The numerical simulations are done directly in the thermodynamic limit with a time-evolution algorithm based on matrix product states.  
The area law of entanglement entropy is observed to hold after the local quench. 
As a result, the simulations can be performed for long times with small bond dimensions. 
Various local quenches on the Ising chain and the three-state Potts chain are studied. 
\end{abstract}
\pacs{Valid PACS appear here}
\maketitle
Physicists have been interested in the apparent paradox between the reversible microscopic laws of dynamics and the irreversible macroscopic phenomena since the time of Boltzmann. 
A basic such phenomenon in quantum systems is dissipation, i.e. the relaxation to the ground state of the Hamiltonian when interacting with an infinitely large external environment.       
Traditionally, dissipation has been studied in the subject of open quantum systems, which approximates the environment as a fluctuating-dissipative effect in the dynamics, such as in the Lindblad master equation \cite{Lindblad, Lindblad_review}. 
In this framework, the dynamics is non-unitary and dissipation occurs naturally. 
However, in principle, dissipation should be explainable in terms of the fundamental unitary quantum dynamics, if one is able to treat the system and the environment together in full detail. 
The quantum dynamics in this case needs to be carried out in the thermodynamic limit, and the evolution time needs to be long enough to reach stationarity.  
In general, this is a daunting task, and a minimalistic physical setting which admits dissipation is desirable. 
In this paper, we provide the phenomenology of such a setting.

Recently, there has been a surge of interest in the quantum dynamics of isolated many-body systems, fueled by the progress in both experimental methods and numerical algorithms.
Most attention has been given to quantum quenches, where one abruptly changes the Hamiltonian so that the quantum state starts as the ground state of the pre-quenched Hamiltonian and evolves thereafter unitarily under the post-quenched Hamiltonian.    
Currently, the only reliable numerical algorithms \cite{itebd, tdvp, iTDVP} to simulate general quantum quenches in the thermodynamic limit are based on matrix product state (MPS) in one dimension (1D) away from quantum criticality.  
To look for dissipation, we consider a local quantum quench in 1D, where the Hamiltonian is only changed in a finite region. 
This finite region will serve as the subsystem of interest, and the rest of the system as the infinite external environment. 
An important motivation to study local quenches has to do with the area law of entanglement entropy \cite{Area_law_review,Area_law_Hastings}, i.e. the fact that, in one dimension, the entanglement entropy of a subsystem does not grow with its size.   
The area law applies to the ground states of non-critical quantum systems.     
The quantum dynamics that gives rise to dissipation should not break the area law, because if the quantum state relaxes to the ground state of the post-quenched Hamiltonian, it should still be an area-law state.    
While extended quantum quenches break the area law in general \cite{Calabrese_2005,Huse_2013}, the local quenches, as we will show, do not.  
Because MPSs can represent area-law states efficiently \cite{MPS_simulability}, this means that one can simulate efficiently the dynamics of local quenches for long times and study the dissipation therein.  
In this paper, we demonstrate that dissipation occurs in non-integrable systems when the local quenching field is of order unity.  
We suggest that a mechanism similar to the eigenstate thermalization hypothesis (ETH) \cite{ETH, ETH_thermalization} is responsible for the dissipation observed.     
When the quenching field is very large, highly oscillatory dynamics is observed and no stationary limit is seen within the time of the simulation.  
We also show incomplete dissipation in the integrable Ising model with its fermionic solution \cite{Kogut}.  

Before we present the results, we briefly discuss the method used to simulate the local quenches, which was developed in \cite{nonuniform_TDVP} and recently improved by us \cite{mixed_iTDVP}. 
In a local quench, the system is composed of three parts: a uniform infinite left bulk, an inhomogeneous finite central part, and a uniform infinite right bulk. 
Because the local quench does not change the Hamiltonian of the two bulks, the bulk MPS tensors stay unchanged. 
The central region is evolved by a finite-size MPS time-evolution algorithm \cite{finite_tdvp}, and takes into account of the influence of the bulks in a simple way described in \cite{mixed_iTDVP}.  
As the information of the local quench spreads, the inhomogeneous region needs to be expanded dynamically with a rate bounded by the Lieb-Robinson velocity \cite{Lieb-Robinson} of the system.
Thus, to evolve a system for time $t$, the computational time scales with $t^2$, as opposed to $\exp(t)$ in an extended quench. 
See \cite{mixed_iTDVP} for further details. 
In the following, $D$ and $\delta t$ denote the bond dimension and the time step of the simulation. 
The simulations below are all done with multiple bond dimensions and the presented results have been converged with $D$.  

The systems we study are the transverse-field Ising chain (TFIC) and the three state quantum Potts chain, both in the paramagnetic phase.  
The Ising chain has Hamiltonian: 
\begin{equation}
  \hat{H}_\Ising = -\sum_{i=-\infty}^\infty \hat{Z}_i \hat{Z}_{i+1} + 1.5 \sum_{i=-\infty}^{\infty} \hat{X}_i 
  \label{eq:Ising}
\end{equation}
where $\hat X, \hat Z$ are the Pauli matrices. 
It is integrable \cite{Kogut}.
The Potts chain has Hamiltonian: 
\begin{equation}
  \hat H_{\Potts} = -\sum_{i=-\infty}^\infty (\hat \tau_i^\dag \hat\tau_{i+1} + \hat \tau_{i+1}^\dag \hat\tau_i) - 1.5\sum_{i=-\infty}^\infty(\hat\sigma_i^\dag + \hat\sigma_{i})
  \label{eq:clock}
\end{equation}
where the operators $\hat \tau_i$ and $\hat\sigma_i$ act on the three states of the local Hilbert space at site $i$, which we label by $\ket{0},\ket{1}$, and $\ket{2}$.  
In this local basis, the $\hat \tau_i$ is a diagonal matrix with diagonal elements $\omega^m$ where $\omega = e^{i2\pi/3}$ and $m = 0, 1, 2$. 
$\hat\sigma_i$ cyclically permutes $\ket{2}_i$ to $\ket{1}_i$, etc., and together with $\hat \sigma_i^\dag$ acts as a transverse-field.   
The Potts chain is non-integrable. 
We denote the pre-quenched Hamiltonian by $\hat H_0$ and the post-quenched Hamiltonian by $\hat H_1$. 
In the following, $\hat H_1 = \hat H_0 + \delta \hat H$, where $\delta \hat H$ is a local field on site $i = 0$. 
We use $\GS0$ and $\GS1$ to denote the ground state of $\hat H_0$ and $\hat H_1$. 

To quantify dissipation, we compute the $k$-body reduced density matrix, $\hat \rho_k(t)$, of $k$ contiguous spin sites centered around site $0$ in the time-evolved state and take its distance from the $k$-body reduced density matrix, $\hat \rho_{k,\GS1}$, in $\ket{\GS1}$:   
\begin{equation}
  d_k(t) \equiv \norm{\hat \rho_k(t) - \hat \rho_{k,\GS1}}_\text{HS}, 
\end{equation}
where $\norm{\cdot}_\text{HS}$ is the Hilbert-Schmidt norm, i.e. $\norm{M}_\text{HS} = \Tr(M M^\dag)$.

\begin{figure}[htb]
  \vspace{3mm}
  \centering
  \caption{Quench I: $\hat H_0 = \hat H_\Ising$ and $\delta \hat H = \hat{X}_0$. 
  $D = 30$ and $\delta t = 0.005$. 
  The inset is a zoom-in of the main plot. 
  }
  \includegraphics[scale=0.45]{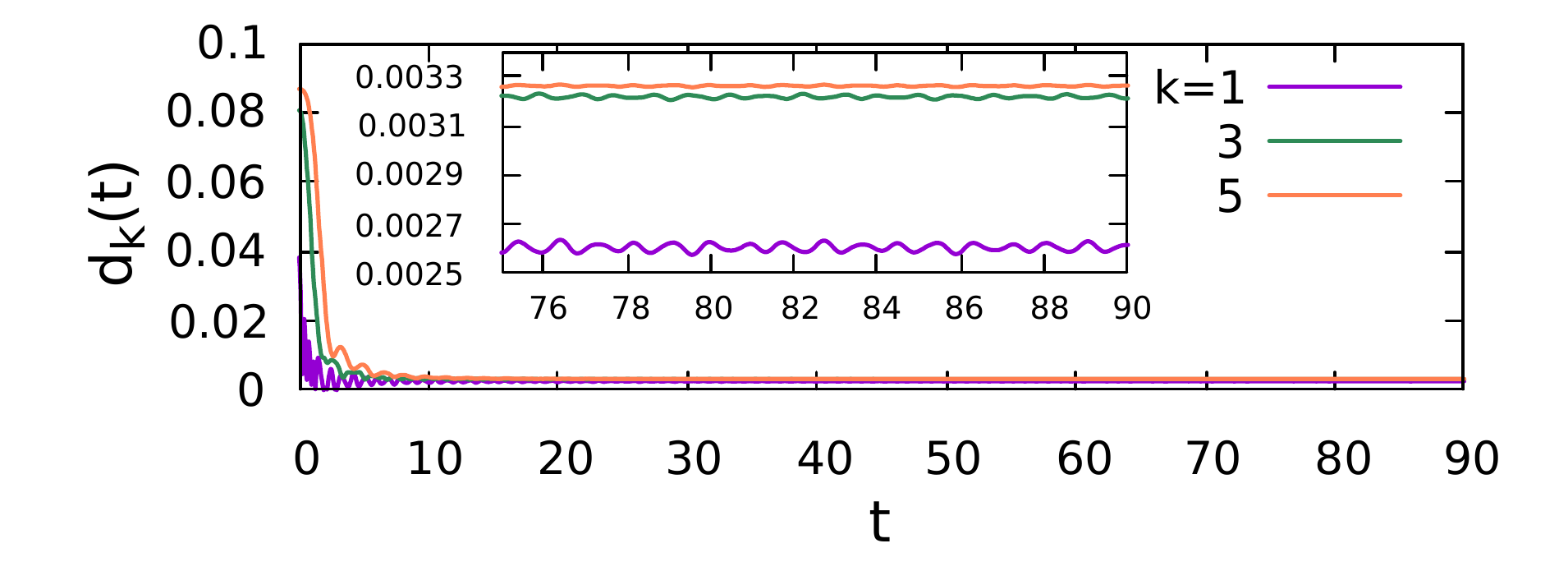}
\label{fig:I}
\end{figure}
We first consider the TFIC. 
We simulate the dynamics after the following quantum quenches: quench I: $\delta \hat H = \hat X_0$, quench II: $\delta \hat H = \hat Z_0$, and quench III: $\delta \hat H = 5 \, \hat Z_0$. 
As shown in Fig. \ref{fig:I}, the quantum state after quench I approaches a stationary limit, close to but different from $\hat\rho_{\GS1}$.   
Because quench I does not break the integrability of $\hat H_0$, we can solve its dynamics completely, presented later. 
The stationary state of this local quench is described by a generalized Gibbs ensemble, $\braket{\cdot}_{\GGE}$.   
For a quench with $\delta \hat H = \delta h_x \hat{X}_0$, $ \braket{\hat O}_\GGE -\braket{\hat O}_{\GS1} \sim (\delta h_x)^2$ for an arbitrary local observable. 
Thus, for a small local quenching field, the dynamics exhibits incomplete dissipation. 

\begin{figure}[htb]
\centering
\caption{Quantum dynamics after the transverse-field Ising chain is quenched with a local longitudinal field.}
\begin{subfigure}{0.5\textwidth}
\caption{ 
Quench II: $\hat H_0 = \hat H_\Ising$ and $\delta \hat H = \hat Z_0$.
$D = 30$ and $\delta t = 0.01$. 
The inset is a zoom-in  of the main plot.
Note the different time ranges on the inset and the main plot.  
}
\includegraphics[scale=0.45]{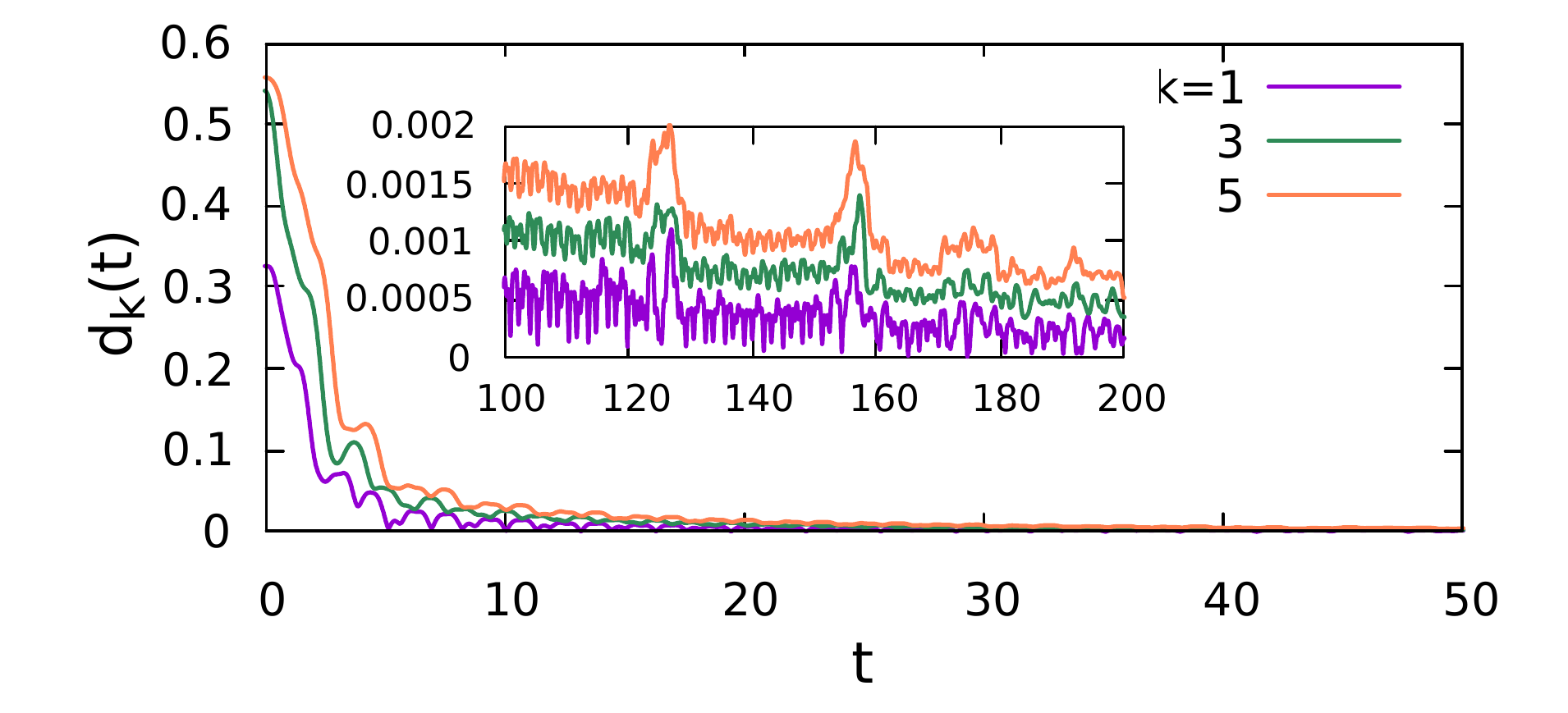}
\end{subfigure}
\begin{subfigure}{0.5\textwidth}
\centering
\caption{
Quench III: $\hat H_0 = \hat H_\Ising$ and $\delta \hat H = 5\,\hat Z_0$.
$D = 20$ and $\delta t = 0.005$. 
The inset shows $\delta X(t) = \braket{\hat X_0(t)} - \braket{\hat X_0}_{\GS1}$.  
}
\includegraphics[scale=0.45]{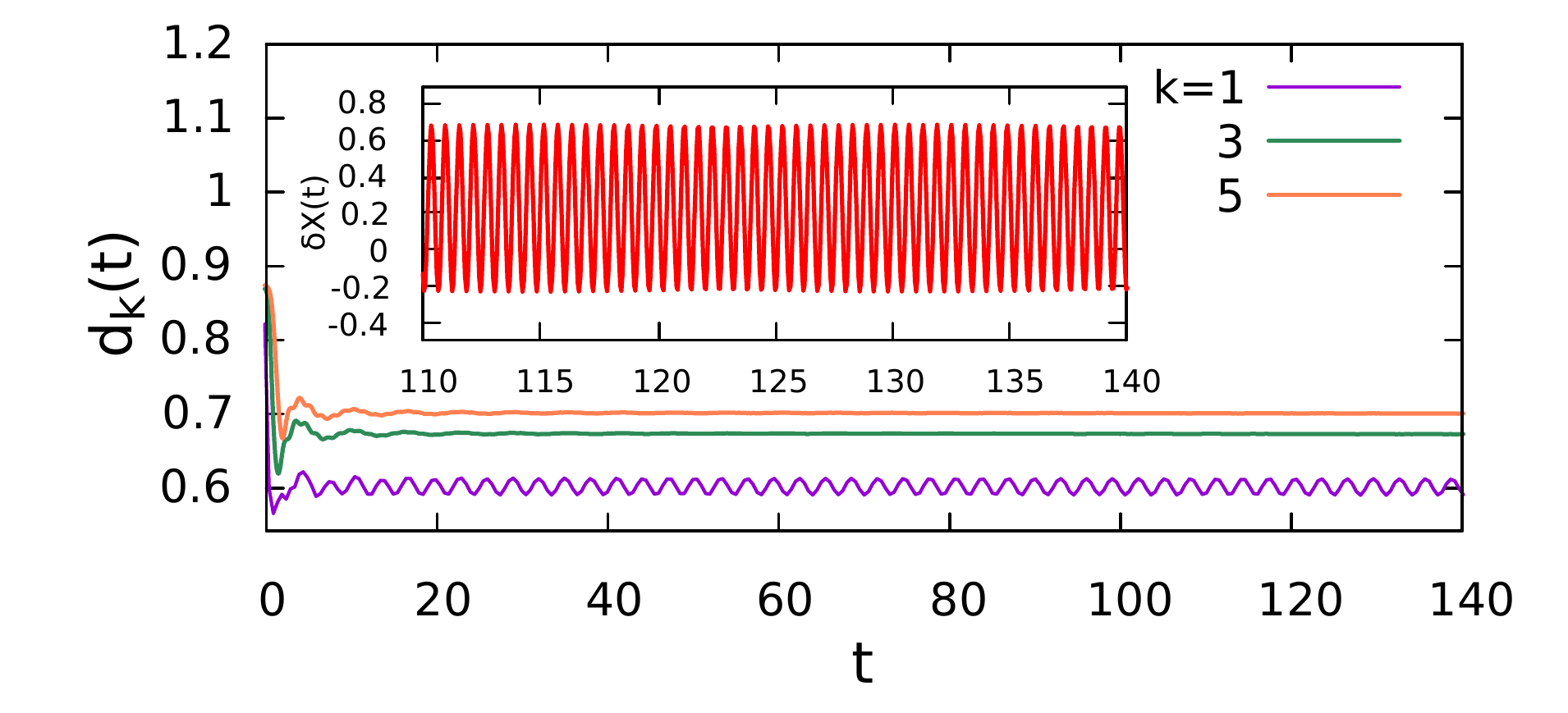}
\end{subfigure}
\label{fig:II}
\end{figure}
\begin{figure}[htb]
\centering
\caption{Quench II: $\delta \hat{H} = \hat{Z}_0$. $D$ = 30 and $\delta t = 0.01$.}
\begin{subfigure}{0.48\textwidth}
\caption{ 
Time-evolution of the half-chain entanglement entropy $S_i(t)$, where the chain is partitioned between site $i$ and $i+1$. 
}
\includegraphics[scale=0.40]{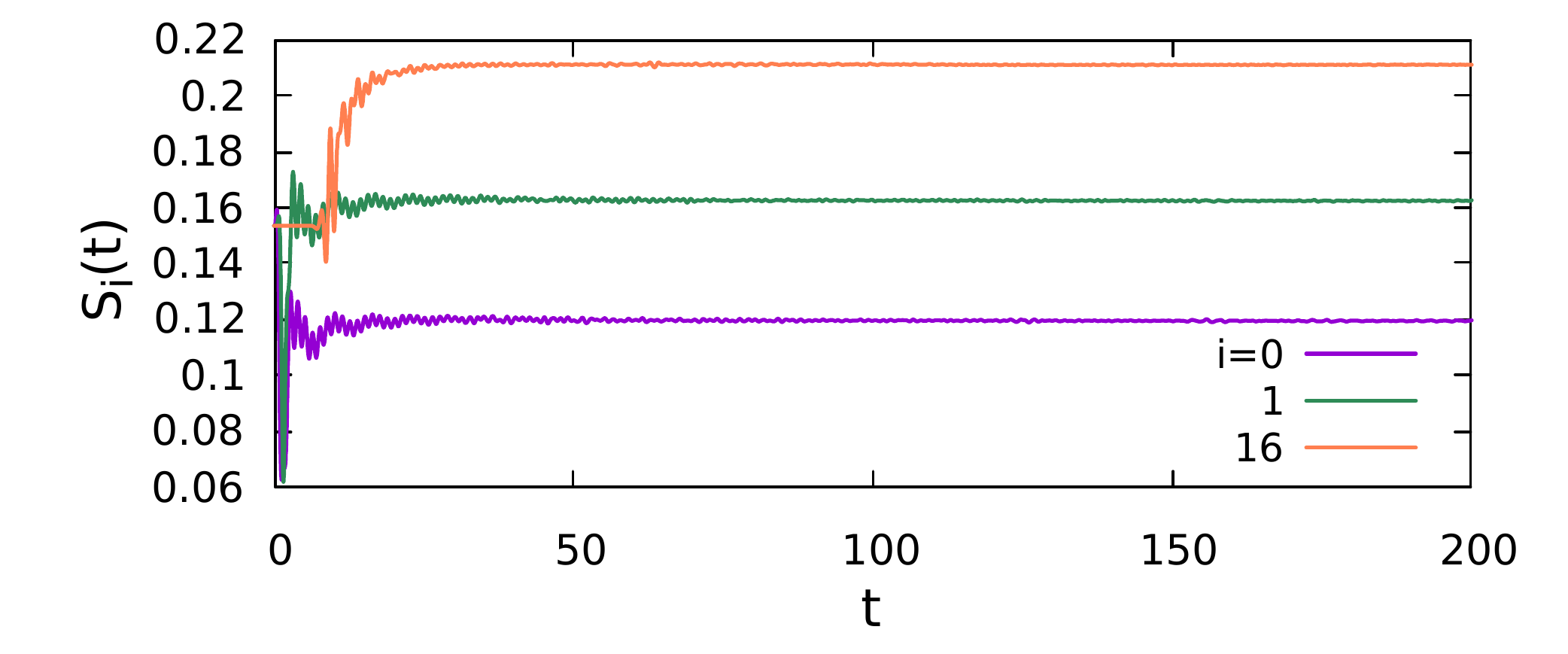}
\end{subfigure}
\begin{subfigure}{0.48\textwidth}
\centering
\caption{
Time-evolution of the transverse magnetization, $\braket{\hat X_i(t)}$, at sites $i = 0, 1,$ and 16. 
The horizontal line under each curve is $\braket{\hat X_i}$ in $\ket{\GS1}$ for each site.
The inset is a zoom-in of the main plot for $i = 0$ and $1$. 
Note the different time ranges. 
}
\includegraphics[scale=0.40]{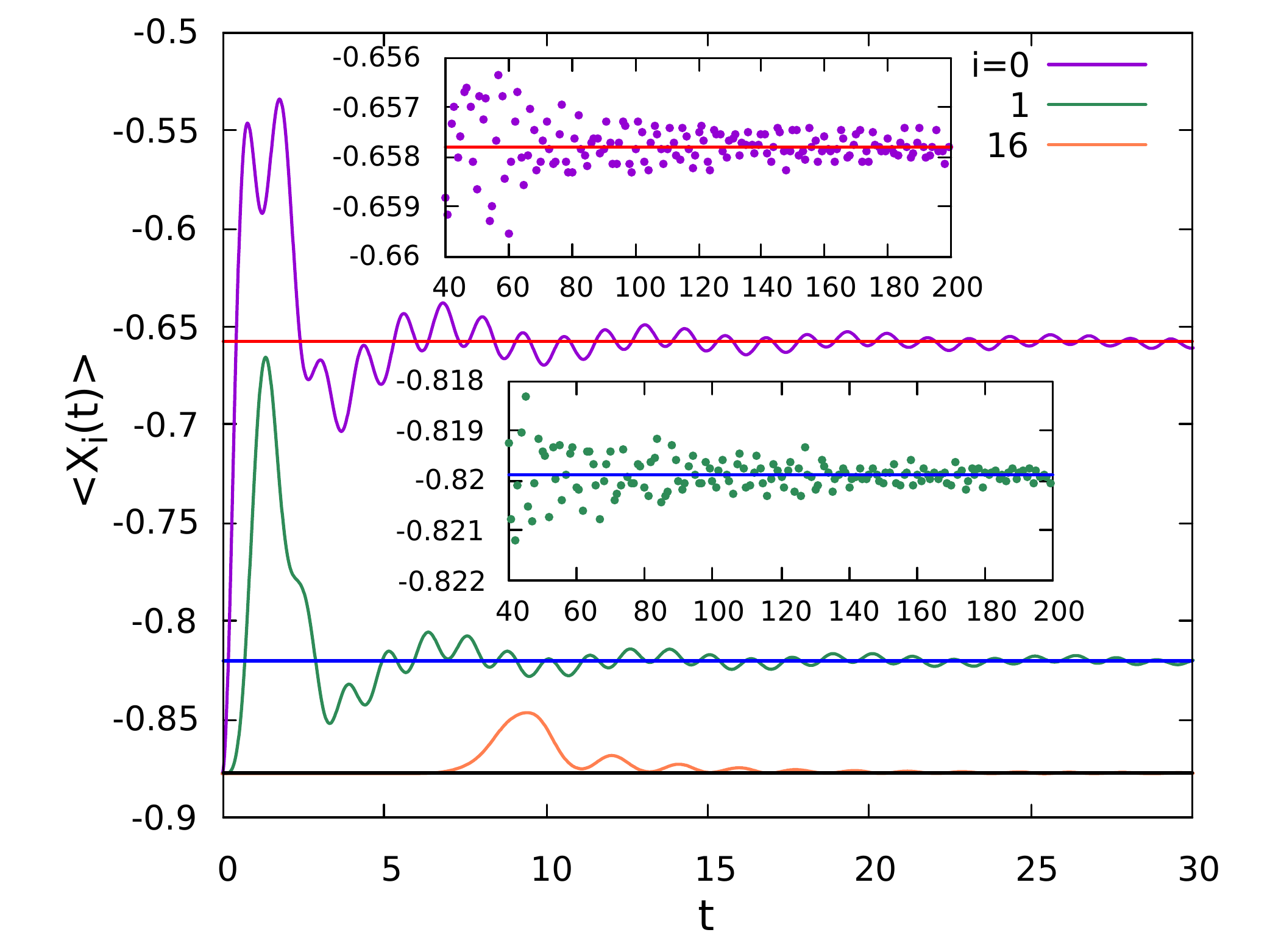}
\end{subfigure}
\label{fig:Hz}
\end{figure}

Quench II and III break the integrability of $\hat H_0$, and the system is no longer integrable.  
As shown in Fig. \ref{fig:II}, when the TFIC is quenched with a longitudinal field, there are two regimes of the long-time behavior of the dynamics.  
When the longitudinal field is of order unity, e.g. in quench II, the system dissipates very fast.
The distance of the one-body reduced density matrices between the time-evolved state and $\ket{\GS1}$ can fall below $10^{-4}$ and shows no sign of stopping decreasing by $t = 200$. 
Fig. \ref{fig:Hz} shows the time-evolution of the half-chain entanglement entropy, $S_i(t)$, and the transverse magnetization, $\braket{\hat X_i(t)}$, of the system at various lattice sites $i$. 
The entanglement entropies saturate to finite values at large time, indicating that the local quench does not break the area law.  
The transverse magnetization relax to their values in $\ket{\GS1}$.    
The amplitude of the oscillation of $\braket{\hat{X}_i(t)}$ also decreases with time.  
In sharp contrast to this dissipative behavior is a local quench where the longitudinal field is large, e.g. in quench III.
As seen in Fig. \ref{fig:II} (bottom) the dynamics in quench III oscillates strongly.  
The amplitude of the oscillation in $\braket{\hat X_0}$ maintains essentially constant from $t = 5$ to $t = 140$ with no sign of decreasing. 
This oscillatory behavior is observed in $\delta \hat H = 10 \, \hat{Z}_0$, too. 
In fact, if an infinitely large longitudinal field is applied in the local quench, one indeed expects an oscillatory, non-stationary long-time limit for the dynamics.  
However, it is not clear how this behavior would persist into large but finite quenching fields.    
While Fig. \ref{fig:II} suggests a phase transition in the dynamics as the quenching longitudinal field is varied, the evidence is only numerical, and a more sophisticated argument will be required to determine whether this transition is genuine, which we leave open here. 

\begin{figure}[htb]
\centering
\caption{Quantum dynamics after the Potts chain is quenched by a local longitudinal field.}
\begin{subfigure}{0.5\textwidth}
\centering
\caption{
Quench IV: $\hat H_0 = \hat H_\Potts$ and $\delta \hat{H} = \ket{0}\bra{0}_0$. 
$D = 20$ and $\delta t = 0.01$. 
The inset is a zoom-in of the main plot. 
}
\includegraphics[scale=0.45]{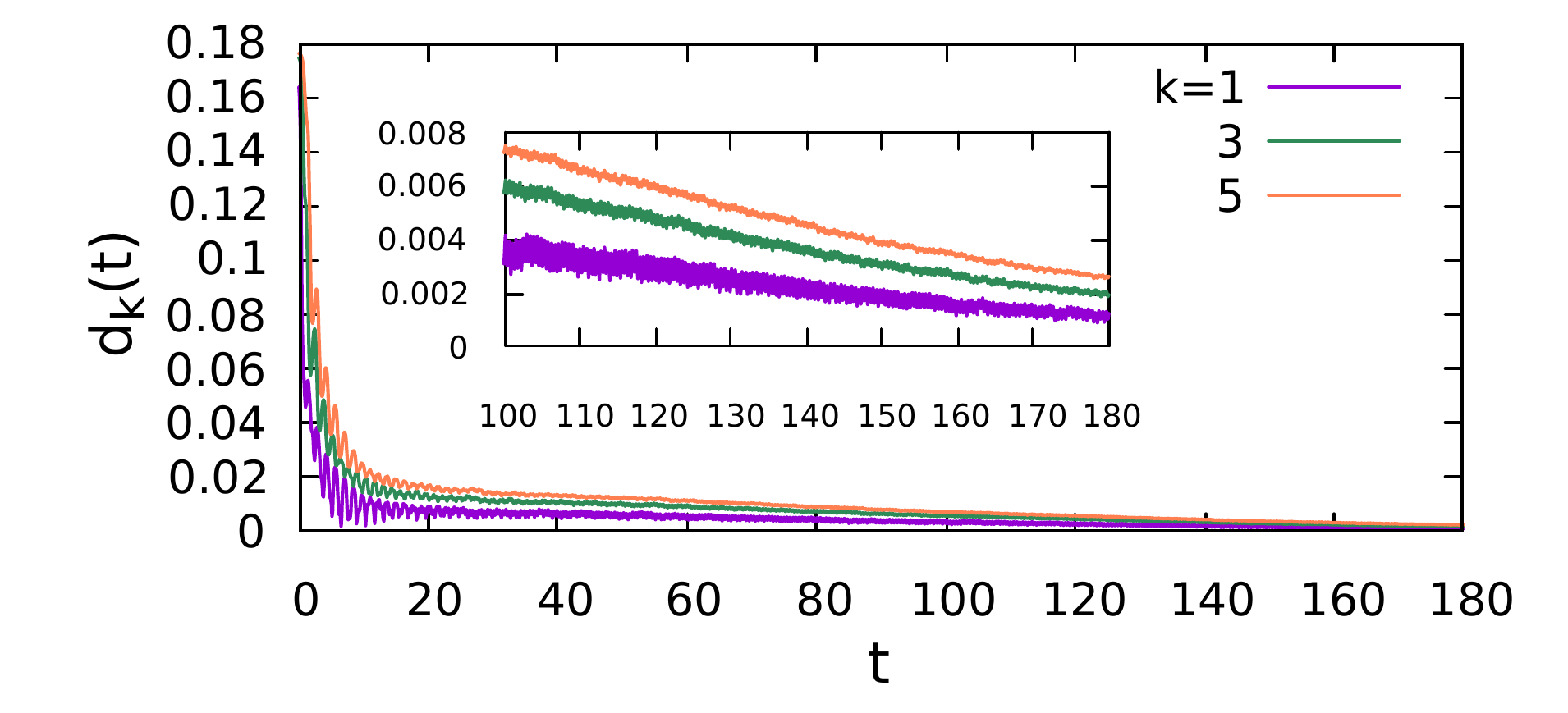}
\end{subfigure}
\begin{subfigure}{0.5\textwidth}
\centering
\caption{Quench V: $\hat H_0 = \hat H_\Potts$ and $\delta \hat{H} = 50\,\ket{0}\bra{0}_0$. 
$D = 20$ and $\delta t = 0.01$. 
The inset shows $\delta O(t) = \braket{\ket{0}\bra{0}(t)} - \braket{\ket{0}\bra{0}}_{\GS1}$ at site 0. 
}
\includegraphics[scale=0.45]{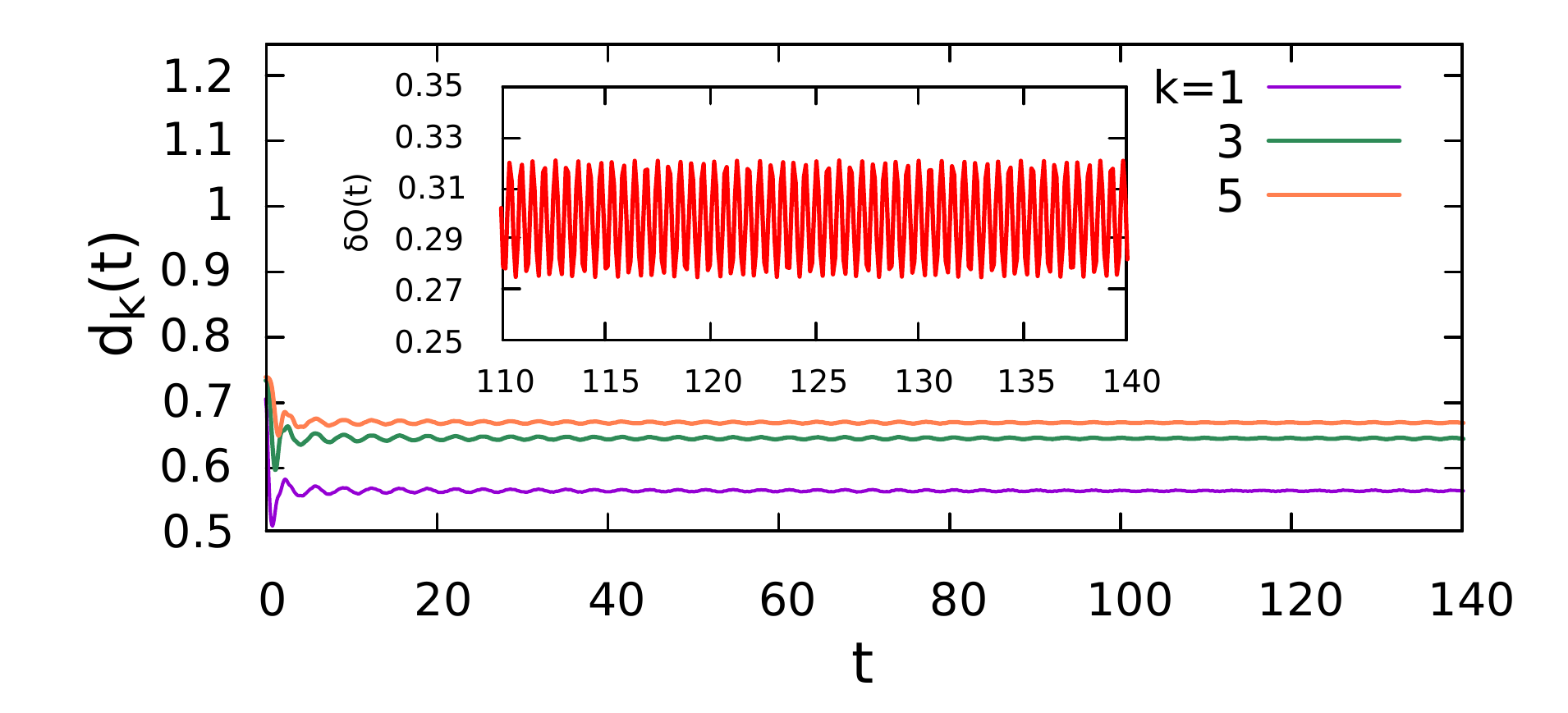}
\end{subfigure}
\label{fig:Potts}
\end{figure}

We now consider $\hat H_0 = \hat H_\Potts$, which is non-integrable. 
We again study the dynamics after a local quench of a longitudinal field on site 0: quench IV: $\delta \hat H = \ket{0}\bra{0}_0$, and quench V: $\delta \hat H = 50\,\ket{0}\bra{0}_0$. 
The long-time behaviors of the dynamics are similar to those observed in the longitudinal quenching of the Ising model.   
As seen in Fig. \ref{fig:Potts}, when the quenching field is of order unity, dissipation of $\hat \rho_k(t)$ to $\hat\rho_{k,\GS1}$ is observed. 
When the quenching field is very large, strongly oscillating dynamics is seen for $t$ up to 140 with no sign of $\hat \rho_k(t)$ approaching $\hat \rho_{k,\GS1}$.    

We now discuss the mechanism for the observed dissipation. 
In quench I, both $\hat H_0$ and $\hat H_1$ are integrable, which allows an exact computation of its quench dynamics.      
Consider the TFIC Hamiltonian of an open chain of size $L$:
\begin{equation}
  \hat H_\TFIC = \sum_{i=1}^{L-1} J_i \hat Z_i\hat Z_{i+1} + \sum_{i=1}^L h_i \hat X_{i}. 
\end{equation}
The TFIC can be mapped into a system of non-interacting fermions, $\{\hat c_i, \hat c^\dag_i\}$ \cite{Kogut, Ising_solution}: $ \hat H_\TFIC = \hat{\vec C}^\dag \Theta \hat{\vec C}$, 
where $\hat{\vec C}$ is a column vector of $2L$ fermion operators with $\hat{\vec C}_i = \hat{c}_i$ and $\hat{\vec C}_{i+L} = \hat{c}_i^\dag$ for $i = 1, \cdots, L$.  
$\Theta$ is a $2L\times 2L$ matrix with the form 
\begin{equation}
  \Theta = \begin{bmatrix} A & B \\ -B & -A \end{bmatrix} = U\Lambda U^\dag = \begin{bmatrix} \alpha & \beta \\ \beta & \alpha \end{bmatrix} \begin{bmatrix} \lambda & 0 \\ 0 & -\lambda \end{bmatrix} \begin{bmatrix} \alpha^\dag & \beta^\dag \\ \beta^\dag & \alpha^\dag \end{bmatrix} 
    \label{eq:diag}
\end{equation}
where $A$ and $B$ are $L\times L$ matrices made of $J_i$ and $h_i$ \cite{Ising_solution}. 
Here $U$ is an orthogonal matrix whose column vectors are the eigenvectors of $\Theta$, and $\Lambda$ is a diagonal matrix whose matrix elements are the eigenvalues of $\Theta$.  
$\alpha$, $\beta$, and $\lambda > 0$ are all $L \times L$ matrices.
This diagonalizes $\hat H_\TFIC$, and gives the time-dependence of the fermion operators under $\hat H_1$: $U_1^\dag \hat{\vec C}(t) = T(t) U_1^\dag \hat{\vec C}$, where $T(t) = \exp(-i 2\Lambda_1 t)$.     
Subscripts $0$ and $1$ on $U$ and $\Lambda$ are respectively for $\hat H_0$ and $\hat H_1$. 
In the TFIC, local observables are determined by the two-body correlators of the fermion operators, e.g. $\braket{\hat{X}_i} = 1-2\braket{\hat c_i^\dag \hat c_i}$.   
Thus, one is interested in the dynamics of the matrix of two-body correlators: 
\begin{equation}
\begin{split}
  &M(t) \equiv \braket{\GS0|\hat{\vec C}(t)\hat{\vec C}^\dag(t)|\GS0}  
  \\
  &= U_1 T(t) U_1^\dag U_0 \braket{\GS0|U_0^\dag \hat{\vec C}\hat{\vec C}^\dag U_0|\GS0} U_0^\dag U_1 T^\dag(t) U_1^\dag. 
\end{split}
\end{equation} 
Note that $U_0^\dag \hat{\vec C}$ are the fermion operators that diagonalize $\hat H_0$, and act on $\ket{\GS0}$ in the standard way.  
Thus, 
\begin{equation}
  E \equiv  \braket{\GS0|U_0^\dag \hat{\vec C}\hat{\vec C}^\dag U_0|\GS0} = \begin{bmatrix} 0 & 0 \\ 0 & I_{L\times L} \end{bmatrix}
\end{equation}
where $I_{L\times L}$ is the $L\times L$ identity matrix. 
We now write $U_1^\dag U_0 = 1 + \Delta$, where $\Delta$ is a small matrix when $\delta \hat H$ is small.  
Because $U_0$ and $U_1$ both have the block structure in Eq. \ref{eq:diag}, so does $\Delta$, which we write as 
\begin{equation}
  \Delta \equiv U_1^\dag U_0 - 1 =  \begin{bmatrix} a & b \\ b & a \end{bmatrix}.
\end{equation}
Then, defining $\Delta(t) \equiv T(t) \Delta T(t)^\dag$, we have 
$M(t)$ = $U_1 E U_1^\dag$ + $\delta M(t)$ 
where $\delta M(t)$ = $U_1 \Delta(t) E U_1^\dag$ + $U_1 E \Delta(t)^\dag U_1^\dag$ + $U_1$ $\Delta(t)E$ $\Delta(t)^\dag U_1^\dag$. 
Note that $U_1EU_1^\dag$ is the correlator matrix in $\ket{\GS1}$, and thus $\delta M(t)$ measures the deviation from complete dissipation. 
Its upper left $L\times L$ block, $\delta M(t)_{\text{UL}}$, can be computed to be    
\begin{equation}
  \begin{split}
    \delta M(t)_{\text{UL}} &= \alpha_1 \Gamma(t) bb^\dag \Gamma(t)^\dag \alpha_1^\dag - \beta_1 \Gamma(t)^\dag bb^\dag\Gamma(t)\beta_1^\dag 
  \\
  &+ [\beta_1 \Gamma(t)^\dag(b^\dag+ab^\dag)\Gamma(t)^\dag\alpha_1^\dag + \text{H.c.}] 
\end{split}
\label{eq:phase}
\end{equation}
where $\Gamma(t) = \exp(-2i\lambda_1 t)$. 
In the thermodynamic limit, for a large $t$, the time-dependent terms in Eq. \ref{eq:phase} will be very out of phase, cancelling one another, and we assume that only the time-independent terms are non-vanishing:  
\begin{equation}
    \lim_{t\rightarrow \infty}(\delta M(t))_{ij} = \sum_{k=1}^L (bb^\dag)_{kk}[(\alpha_1)_{ik}(\alpha_1^\dag)_{kj} 
- (\beta_1)_{ik}(\beta_1^\dag)_{kj}].
\label{eq:DE}
\end{equation}
Eq. \ref{eq:DE} works very well.  
For example, in the dynamics of Fig. \ref{fig:I}, $\braket{\hat X_0(t)} = -0.92776 \pm 0.00003$ at $t = 90$, whereas Eq. \ref{eq:DE} gives -0.927762 for $L = 1024$.  
Assuming $b$ is to the linear order of $\delta h_x$, we conclude that the deviation from complete dissipation is second order in $\delta h_x$.  

For quench II and IV, we no longer have the luxury of integrability, and numerical means must be employed. 
We expand the mean value of a local observable $\hat O$ into time-dependent and time-independent terms: 
\begin{equation}
  \braket{\hat O(t)} = \sum_{n,m}^{n\not= m} e^{-i(E_m-E_n)t}c^*_m c_n \braket{m|\hat O|n} + \sum_{n} |c_n|^2\braket{n|\hat O|n}, 
\end{equation}
where $n, m$ are the eigenstates of $\hat H_1$, and $c_n = \braket{n|\GS0}$. 
We assume that for large $t$, the time-dependent terms cancel so that their sum vanishes, and that the stationary values of $\braket{\hat O(t)}$ is given by the diagonal ensemble \cite{ETH_thermalization}, $\hat\rho_\DE$:    
\begin{equation}
  \lim_{t\rightarrow \infty}\braket{\hat O(t)} = \Tr(\hat \rho_\DE \hat O), \hspace{3mm} \hat\rho_\DE = \sum_{n} |c_n|^2 \ket{n}\bra{n}.
  \label{eq:O_DE}
\end{equation}
The employment of such a diagonal ensemble is a key step in the understanding of thermalization with the ETH \cite{ETH_thermalization}. 
Note that Eq. \ref{eq:O_DE} is not a trivial assumption. 
For example, quench III does not seem to satisfy it. 
\begin{figure}[htb]
\centering
\caption{The transverse magnetization, $\braket{\hat X_i}$ in the first 40 energy eigenstates of $\hat H_1 = \hat H_\Ising + \hat Z_0$ on an Ising chain with 101 chains. 
The magnetization profile is shown for the selected energy states $n = 0, 1, 5, 10, 20, $ and 39. 
It is symmetric around $i = 0$, so only $\braket{\hat X_i}$ on $i \ge 0$ is shown. 
The inset shows the transverse magnetization on the zeroth site, $\braket{\hat X_0}$, for all the energy eigenstates $n = 0$ to $39$. 
}
\includegraphics[scale=0.35]{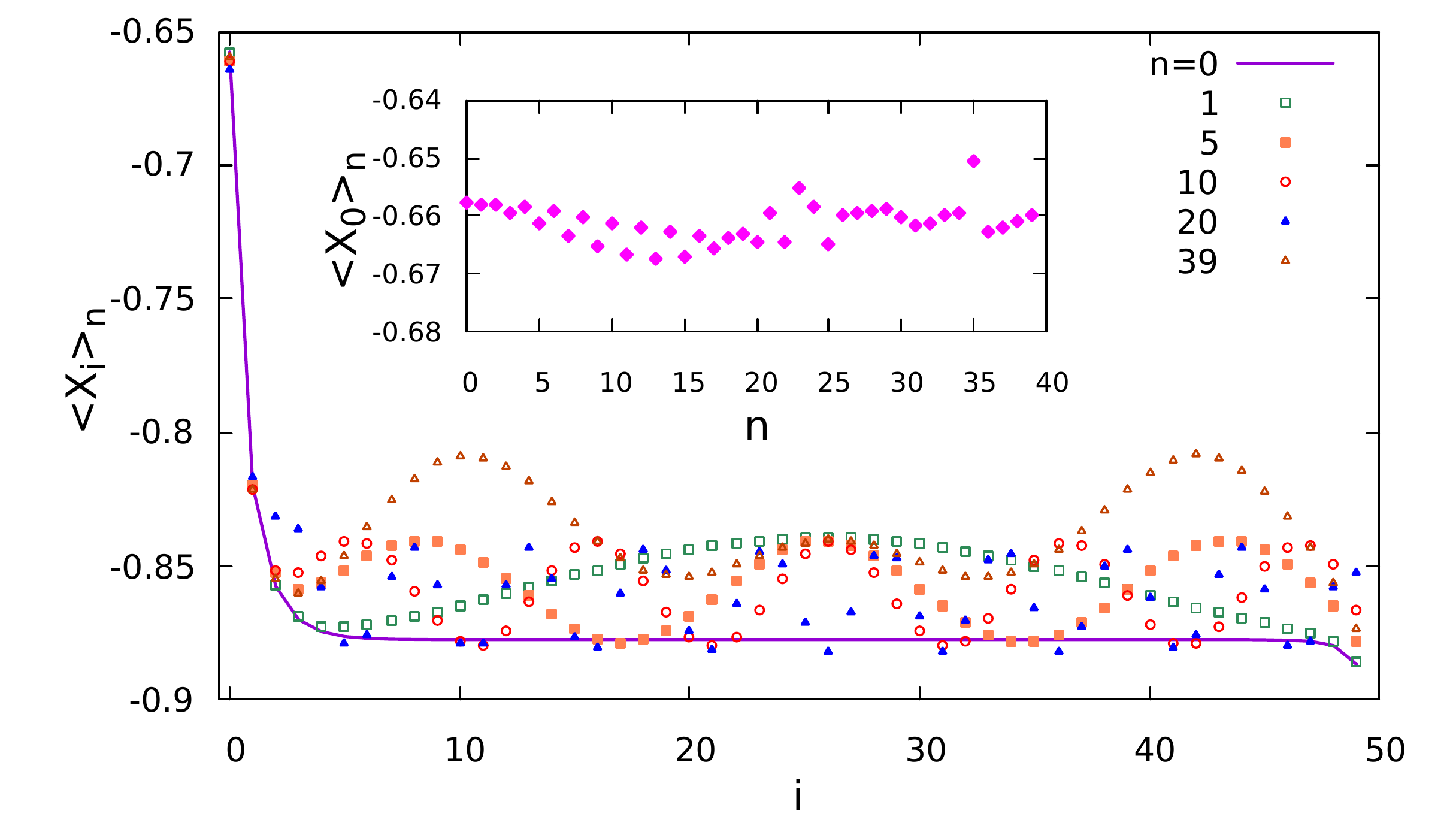}
\label{fig:excite}
\end{figure}
However, inspired by the success of Eq. \ref{eq:DE} and the ETH, we cavalierly proceed to check how well Eq. \ref{eq:O_DE} explains the observed dissipation in quench II.  
Let $\ket{n}$ denote the $n$th energy eigenstate of $\hat H_1$.
Because the quench is local, we assume that $|c_n|^2$ is only significant if $\ket{n}$ does not have extensively more energy than $\ket{0}$, i.e. we consider {\it low-lying} excited states.  
We consider an open Ising chain with $L=101$ sites, and compute the first $40$ energy eigenstates of $\hat H_1$ with density matrix renormalization group \cite{DMRG, White_DMRG}.
These states are converged to have inner products among one another on the order of $10^{-8}$.   
In quench II, $c_0 = \braket{0|\GS0} = \braket{\GS1|\GS0} = 0.9063$. 
Thus, $|c_0|^2\ket{0}\bra{0}$ alone accounts for 82\% of $\hat \rho_\DE$, and is entirely the same with $\hat\rho_{\GS1}$.  
The rest comes from the excited states of $\hat H_1$. 
For this $L=101$ chain, the first 40 energy eigenstates of $\hat H_1$ make up 92\% of the pre-quenched state: $\sum_{n=0}^{39} |\braket{n|\GS0}|^2 = 0.92$. 
In Fig. \ref{fig:excite}, one sees that, {\it locally}, observables in the excited states of $\hat{H}_1$ are very close to those in $\ket{\GS1}$.   
The mean of $\braket{\hat X_0}$ in the first 40 energy states is -0.661 with a standard deviation of 0.003, to be compared with $\braket{\hat X_0}_{\GS1} = -0.658$. 
For $\braket{\hat X_1}$, it is $-0.815 \pm 0.009$, while $\braket{\hat X_1}_{\GS1} = -0.820$. 
This is reminiscent of the ETH \cite{ETH_thermalization}, where local observables in all the energy eigenstates with the same energy have the same value.  
Here, however, the ETH-like phenomenon only occurs for observables confined to the vicinity of site 0.  
As seen in Fig. \ref{fig:excite}, magnetization away from site 0 have very different values in different energy eigenstates. 
Thus, this local ETH-like mechanism only explains the long-time behavior of the sites in the vicinity of site 0.  
A more comprehensive analysis of it will be necessary and is left for future work. 

In this paper, we provided a rich phenomenology of long-time quantum dynamics after a local quench, including dissipation and strong oscillation. 
Two questions are raised. 
One is whether there is a genuine transition between the oscillatory dynamics and the dynamics which has a stationary limit.  
The other one concerns with the nature of the local ETH-like mechanism described above. 
In the end, we ask a third one: in the hydrodynamics approach, e.g. \cite{Transport}, to quantum dynamics, does a dissipative term emerge generically in the transport equation following a local quench?  
\begin{acknowledgments}
  The code is based on ITensor \cite{ITensor} (version 3, C\texttt{++}), and is available upon request. 
The author is grateful for mentorship from his advisor Roberto Car at Princeton. 
He acknowledges support from the DOE Award DE-SC0017865. 
\end{acknowledgments}

\bibliography{abc}
\end{document}